\makeatletter
\declare@file@substitution{revtex4-1.cls}{revtex4-2.cls}
\makeatother

\documentclass[twocolumn]{aastex631}

\usepackage{amsmath}
\newcommand{\msun}{M$_{\odot}$}
\newcommand{\lsun}{L$_{\odot}$}

\received{}
\revised{}
\accepted{}
\submitjournal{The Astrophysical Journal}
\shorttitle{Neutrinos from Carbon-Burning RSGs}
\shortauthors{Seong et al.}

\begin{document}

\title{Neutrinos from Carbon-Burning Red Supergiants and Their Detectability}

\author[0009-0004-9886-3249]{Gwangeon Seong}
\affiliation{Department of Physics, College of Natural Sciences, UNIST, Ulsan 44919, Korea}
\author[0000-0002-2304-7798]{Kyujin Kwak}
\affiliation{Department of Physics, College of Natural Sciences, UNIST, Ulsan 44919, Korea}
\author[0000-0002-5455-2957]{Dongsu Ryu}
\affiliation{Department of Physics, College of Natural Sciences, UNIST, Ulsan 44919, Korea}
\affiliation{Korea Astronomy and Space Science Institute, Daejeon 34055, Korea}
\author[0009-0002-3265-0502]{Bok-Kyun Shin}
\affiliation{Department of Physics, College of Natural Sciences, UNIST, Ulsan 44919, Korea}
\affiliation{Pohang Accelerator Laboratory, Pohang-si, Gyeongsangbuk-do, 37673, Korea}

\correspondingauthor{Dongsu Ryu}\email{dsryu@unist.ac.kr}
\correspondingauthor{Kyujin Kwak}\email{kkwak@unist.ac.kr}

\begin{abstract}

Stars emit MeV neutrinos during their evolution via nuclear syntheses and thermal processes, and detecting them could provide insights into stellar structure beyond what is accessible through electromagnetic wave observations. So far, MeV neutrinos have been observed from the Sun and SN 1987A. It has been suggested that pre-supernova stars in the oxygen and silicon burning stages would emit enough MeV neutrinos to be detectable on Earth, provided they are in the local universe. In this study, we investigate the prospect of detecting neutrinos from red supergiants (RSGs) in the carbon-burning phase. In our Galaxy, around a thousand RSGs have been cataloged, and several are expected to be in the carbon-burning phase. We first calculate the luminosity and energy spectrum of neutrinos emitted during the post-main-sequence evolution of massive stars. For a nearby carbon-burning RSG located $\sim200$ pc away, we estimate the neutrino flux reaching Earth to be as large as $\sim10^5$ cm$^{-2}$s$^{-1}$ with a spectrum peaking $\sim0.6$ MeV. We then assess the feasibility of detecting these neutrinos in underground facilities, particularly in hybrid detectors equipped with water-based liquid scintillator and ultra-fast photodetectors. In detectors with a volume comparable to Super-Kamiokande, for the above flux, we anticipate up to $\sim50$ neutrino events per year with directional information. Although this is a fair number, the number of events from radioactive backgrounds would be much larger. Our results indicate that studying neutrinos from carbon-burning RSGs and predicting supernovae well in advance before their explosion would be challenging with currently available detector technologies.

\end{abstract}


\section{Introduction} \label{sec:intro}

\begin{table*}[t]
\caption{Catalog of red supergiants within 1 kpc}
\begin{tabular}{cccccc}
\toprule
Alias & SIMBAD ID  & Distance {[}pc{]}  & T$_{\rm{eff}}$ {[}K{]} & Luminosity {[}\lsun{]}  & Mass {[}\msun{]} \\
\hline
Betelgeuse  & alf Ori & $168_{-15}^{+27}$  & $3600\pm200$  & $126000_{- 50000}^{+ 83000}$ & 16.5$\sim$19  \\
Antares & alf Sco  & 170  & $3660\pm200$  & $98000_{-29000}^{+40000}$ & 11$\sim$14.3  \\
5 Lacertae  & 5 Lac  & $505.05$ & $3660\pm200$ & $17473\pm3344$ & $5.11\pm0.18$ \\
119 Tauri & 119 Tau  & 550  & $3820\pm135$  & $66000_{-20000}^{+21000}$ & 14.37$_{-2.77}^{+2.00}$ \\
NO Aurigae  & NO Aur  & 600  & 3700  & 67000 & - \\
V424 Lacertae & V424 Lac   & 623  & $3790\pm110.5$  & 11176.69 & - \\
KQ Puppis & KQ Pup   & 659  & $3660\pm170$  & 59800 & 13$\sim$20  \\
MZ Puppis & MZ Pup   & 703  & $3745\pm170$  & 19586.643  & - \\
$\mathrm{\mu}$ Cephei  & mu Cep    & 940$_{-40}^{+140}$ & $3551\pm136$  & 269000$_{-40,000}^{+111,000}$ & 15$\sim$20  \\
V419 Cephei & V419 Cep   & 941  & $3660\pm170$  & 17693.234  & - \\
\hline
\end{tabular}
\label{tab:catalog}
\end{table*}

Astronomy is an observational science that has advanced through the advent of new observational technology. Traditionally, astronomers have relied on detecting electromagnetic waves such as visible light and radio waves from celestial objects and events. However, a new era of astronomy has recently emerged with facilities that allow astronomers to detect non-electromagnetic-wave signals. This new field, called multi-messenger astronomy, encompasses observations of cosmic rays, gravitational waves, and neutrinos.

Among these, astronomy with neutrinos has a unique history. The first detection of astronomical neutrinos was made in the 1960s through the Homestake experiment, which successfully identified neutrinos emitted from the Sun \citep{HomestakeExp}. Since then, solar neutrinos have been measured using a variety of detectors, including Kamiokande \citep{SK1989}, Sudbury Neutrino Observatory \citep{SNOExp}, and Borexino \citep{BorexinoPP, BorexinoCNO}. Neutrinos from the supernova SN 1987A, which exploded in the Large Magellanic Cloud in 1987, were also detected at Kamiokande \citep{Hirata1987}, IMB \citep{Bionta1987}, and Baksan \citep{Alekseev1987}.

Most astronomical neutrino facilities currently in operation are water Cherenkov detectors (WCDs), which detect photons generated when incoming neutrinos interact with electrons or nuclei in water. These facilities can be categorized into two types based primarily on size and corresponding energy windows: large-volume WCDs that are optimized for neutrinos with energies above TeV and relatively small underground WCDs that register mostly neutrinos below TeV. Examples of the first type are IceCube \citep{IceCube2006}, KM3NET \citep{2006Katz_KM3NeT}, and Baikal-GVD \citep{2024Aynutdinov_Baikal}. The second type includes three generations of Kamiokande detectors: Kamiokande \citep{Kaminokande1984}, Super-Kamiokande \citep{SK2003_Fukuda}, and Hyper-Kamiokande \citep{HyperK2018}. High-energy neutrinos above TeV are produced mostly through collisions between cosmic rays and background protons and photons, typically in extreme environments involving shocks, jets, and accretion disks. In contrast, lower-energy neutrinos below GeV are mainly generated via nuclear reactions, such as beta decay, or via various thermal processes. Hence, the two types of neutrino detectors target different astronomical events and phenomena.

Despite historical achievements and ongoing efforts, the number of confirmed astronomical neutrino sources remains limited. To date, besides the Sun and SN 1987A, the only confirmed source is TXS 0506+056, a blazar identified by IceCube as emitting TeV and PeV neutrinos \citep{Aartsen2018}. For this reason, there is a need to discover or identify additional sources to elevate neutrino astronomy as a main component of multi-messenger astronomy. Additionally, theoretical efforts to predict potential astronomical sources and estimate neutrino fluxes from them should accompany observational initiatives.

Promising sources, especially for low-energy neutrinos, include massive stars around the end of their lives. The detection of neutrinos from SN 1987A demonstrated that the death of a massive star releases a huge amount of neutrinos, accounting for most of the gravitational energy released during the collapse of an iron core. If supernovae (SNe) occur in the local universe, the neutrinos produced should be detected by current facilities, such as those listed above, providing insights into the mechanisms of explosions and the physical conditions of their progenitor stars. However, the major limitation with SNe is their low explosion rate within the detectable distance range of neutrinos. For example, in our Galaxy, the last recorded supernova, SN 1604 (Kepler's Supernova), occurred in 1604, although the remnant G1.9+0.3 suggests the possibility of a more recent SN in the late 19th century \citep{Reynolds_2008}. The rate of SN explosions in our Galaxy is estimated to be approximately one or two per century \citep{Tammann1994_snrate, Rozwadowska2021_snrate}. This rarity is the reason why, since SN 1987A, no neutrinos have been detected from SNe for more than three decades.

In addition to SN explosions, it has been shown that massive stars at the final stages of oxygen (O) and silicon (Si) burning can also emit detectable amounts of neutrinos, particularly in underground WCDs, if they are located in our Galaxy and the Local Group \citep{Kato2015, Kato2017, Patton2017a, Patton2017b}. However, the duration of O and Si burning is relatively short, lasting only months and days to hours, respectively, before SN explosions. Hence, the expected number of massive stars in the O and the Si-burning stages would be small, likely much less than one in our Galaxy.

\begin{figure*}[t]
\vskip -0.3 cm
\hskip 2 cm
\includegraphics[width=0.8\textwidth]{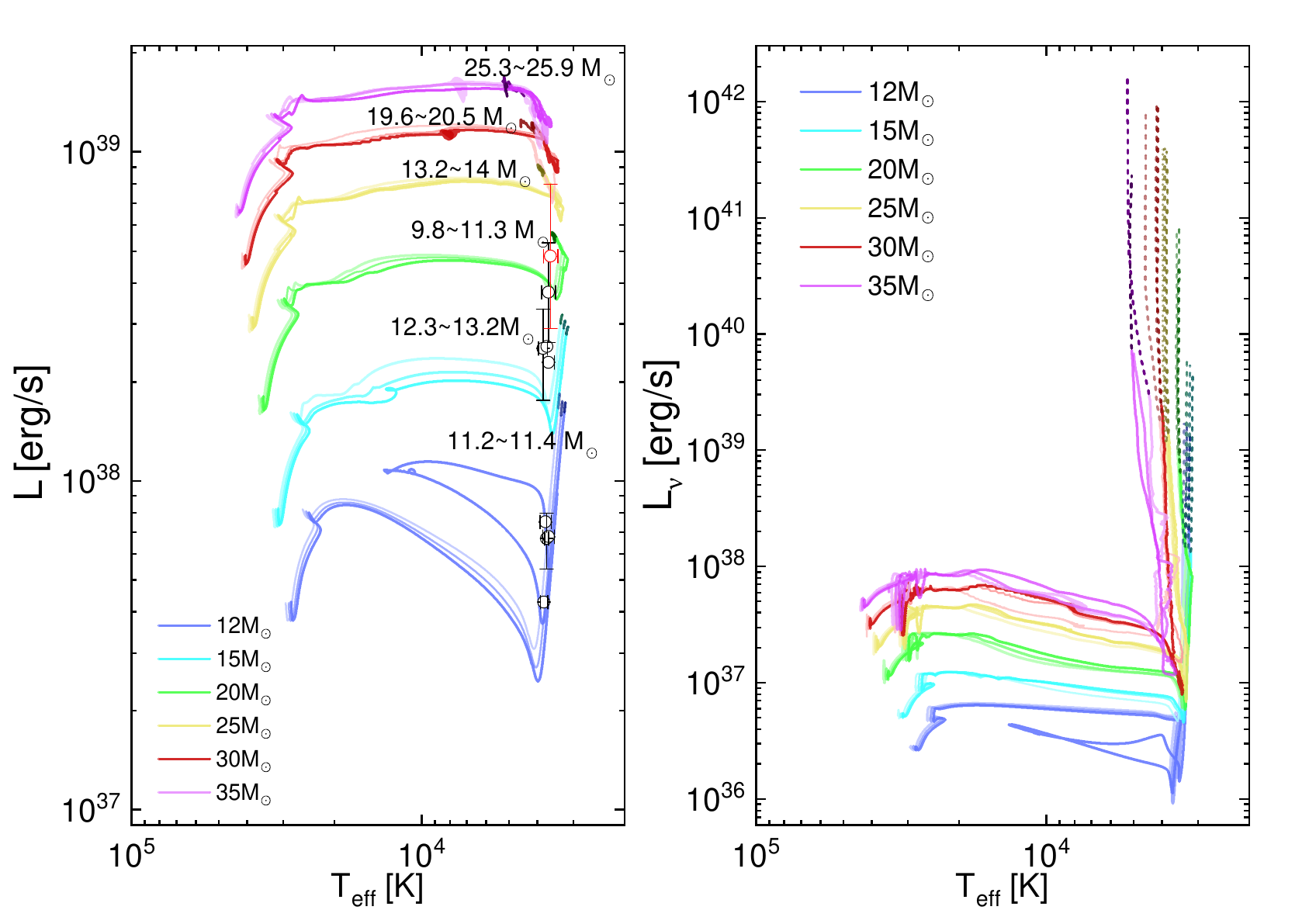}
\vskip -0.2 cm
\caption{Photon (left) and neutrino (right) HR diagram illustrating the evolution of 18 model stars from ZAMS to the end of C-burning. Varying colors indicate stars of different masses, while stars with three different metallicities at ZAMS are presented in dark ($Z=0.02$), intermediate ($Z=0.015$), and light ($Z=0.01$) colors. The C-burning phase is highlighted with dark lines in the left panel and dark dashed lines in the right panel. The left panel also displays the ranges of final masses for stars with different masses at ZAMS. Additionally, on the left panel, the nearest RSG, Betelgeuse, is marked with a red circle and error bars, and nearby RSGs within 1 kpc listed in Table \ref{tab:catalog} are marked with black circles and error bars. See Table \ref{tab:catalog} for the details of these RSGs.}
\label{fig:TwoHR}
\end{figure*}

In contrast to O and Si-burning, the carbon (C) burning phase lasts substantially longer during the late evolution of massive stars. This means that the chance for the presence of C-burning massive stars in our Galaxy should be higher. Such massive stars are expected to be observed as Red Supergiants (RSGs). Approximately a thousand RSGs have been cataloged in our Galaxy \citep{Messineo2019, Healy2024}, with ten within the neighborhood of 1 kpc, including the closest ones, Betelgeuse and Antares located about 170 pc away (see Table \ref{tab:catalog}). Most RSGs should be in the helium (He) burning phase. Considering that the duration of C-burning is approximately a thousand times shorter than that of He-burning \citep{Woosley2002}, around one of the cataloged RSGs might be in the C-burning phase. On the other hand, C-burning lasts a few hundred to a thousand years \citep{Woosley2002}. Hence, given that some fractions of SNe are the core-collapse type, it is likely that several C-burning RSGs exist in our Galaxy. Yet, the chance that one of the nearby RSGs within 1 kpc is currently in the C-burning phase would be small, perhaps at $\sim1\%$ or so. Despite these numbers, it is known that RSGs in the C-burning phase emit a significant amount of neutrinos \citep{Farag2020, Farag2024}, raising the question of whether these neutrinos could be detected. In this paper, we aim to address this question, specifically investigating the feasibility of detecting such neutrinos using hybrid detectors equipped with water-based liquid scintillator and ultra-fast photodetectors.

The paper is organized as follows. In the next section, we describe our model stars that evolve through the end of C-burning. We then present the luminosity of emitted neutrinos, as well as their energy spectrum, during the C-burning phase in RSGs. We also present the neutrino flux as a function of energy that reaches Earth from nearby RSGs, accounting for neutrino oscillation. In Section \ref{sec3}, we explore the detectability of these neutrinos at terrestrial detectors. Considering an ideal setup of hybrid detectors, we present an estimation of the event rate with the nearby RSG neutrino flux.  We also briefly discuss the background noises that must be considered in real detectors. Finally, in Section \ref{sec:discussion}, we summarize our findings and their implications.

\begin{figure*}[t]
\vskip 0 cm
\hskip 0.3 cm
\includegraphics[width=0.94\textwidth]{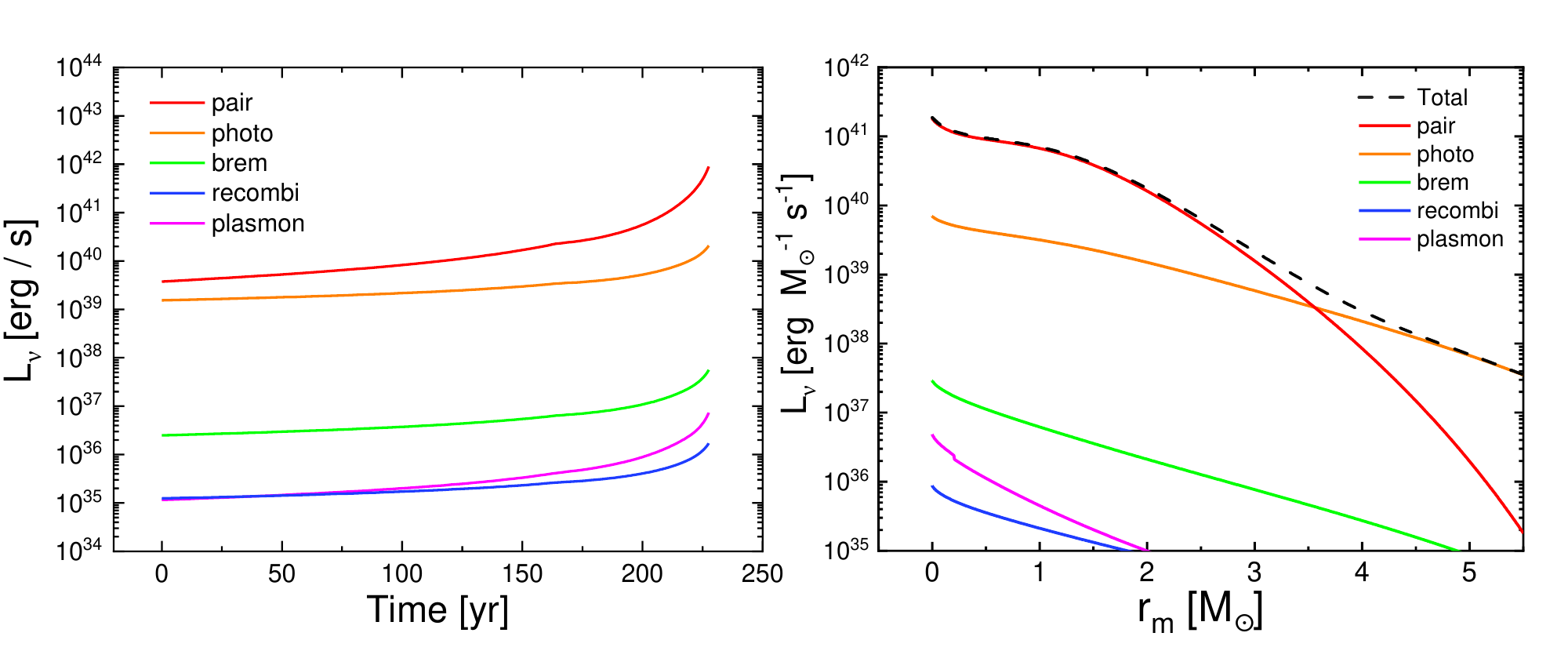}
\vskip -0.2 cm
\caption{Left: Time evolution of neutrino luminosities resulting from various thermal processes during the C-burning phase for our reference star (30 \msun~and $Z=0.02$ at ZAMS). Here, the beginning of C-burning is set to year zero. Right: Radial profile of neutrino luminosities for the same reference star at 210 years after the beginning of C-burning. Here, the $x$-axis represents the mass coordinate, and the $y$-axis shows the luminosity per mass. The model has a central temperature of $\log{T_{c}}=8.96$ ($T_{c}$ in units of K) and a central electron density of $\log{\rho Y_{e}}=4.92$ ($\rho Y_{e}$ in units of g cm$^{-3}$) at $r_m=0$.}
\label{fig:neutrinolum}
\end{figure*}

\section{Neutrino Flux from RSGs}

To generate a sample of model stars, we utilize the Modules for Experiments in Stellar Astrophysics \citep[{\tt MESA}, revision r12115;][]{Paxton2011, Paxton2013, Paxton2015, Paxton2018, Paxton2019, Jermyn2023}. We follow the evolutionary track for 18 stars with initial masses $M =$ 12, 15, 20, 25, 30, and 35 \msun~and initial metallicity $Z =$ 0.02, 0.015, and 0.01, starting from the zero-age main sequence (ZAMS) up to the end of C-burning. Here, the beginning of C-burning is defined as the point when $\sim 1$ \% of core $^{12}$C, produced by He-burning, is consumed, and the end of C-burning is marked by the core $^{12}$C mass fraction dropping to $\sim 1$ \%. We designate the 30 \msun~star with solar metallicity, $Z =$ 0.02, as the reference star and provide detailed analysis for this star as needed.

Our models include single (not binary), non-rotating, and mass-losing stars. We incorporate mass loss using the {\tt Dutch} scheme with a scaling factor of 0.8, which combines the prescriptions from \citet{Vink2001} for hot stars and \citet{deJager1988} for cool stars. To account for convection, we set the mixing length parameter to $\rm{\alpha_{mlt}}=1.5$; additionally, we employ the Ledoux criterion for the convective boundary, with a semi-convection parameter $\alpha_{sc}=0.01$. For nuclear reactions, we use the large built-in reaction network {\tt mesa\_204}, which is based on reaction rates from the REACLIB library \citep[the current standard version is dated as of October 20, 2017; ][]{Cyburt2010}. Other model parameters are set to those in the {\tt MESA\_inlist} of \citet{Farag2020}\footnote{This {\tt inlist} file is available at\dataset[doi:10.5281/zenodo.3634068]{https://doi.org/10.5281/zenodo.3634068}.}. The files to generate our model stars are available on Zenodo under an open-source 
Creative Commons Attribution license:\dataset[doi:10.5281/zenodo.14791656]{https://doi.org/10.5281/zenodo.14791656}.

In stars, neutrinos are produced through two kinds of processes during their evolution: nuclear reactions and thermal processes. To compute neutrino production via nuclear reactions, predominantly $\beta^{\pm}$-decay and electron capture, we utilize the default reaction rates of {\tt MESA}, which are derived from the tables of \citet{Langanke2000}, \citet{Oda1994}, and \citet{Fuller1985}; the neutrino energy emissivity, $Q$, in units of energy per volume per time, is calculated using the reaction rates. For thermal neutrino production, we employ the thermal neutrino module in {\tt MESA}, which includes all thermal processes, electron-positron pair annihilation, photo-neutrino production, bremsstrahlung, plasmon decay, and recombination; the module calculates $Q$ for thermal neutrinos using the fitting formulae of \citet{Itoh1996a, Itoh1996b}. The neutrino luminosity, $L_{\nu}$, in units of energy per time, is then obtained by integrating $Q$ over the entire volume of the star.

Figure \ref{fig:TwoHR} presents the conventional Hertzsprung-Russell (HR) diagram on the left panel and the neutrino HR diagram on the right panel for our 18 model stars. The C-burning phase is highlighted by dark lines in the left panel and dark dashed lines in the right panel, near the end of evolutionary tracks. In the left panel, the ranges of final masses are given; for example, three model stars with 30 \msun~and $Z=0.02$, 0.015, and 0.01 at ZAMS (reddish lines) have the final masses ranging from 19.6 \msun~to 20.5 \msun~when they evolve to the end of C-burning.

In the neutrino HR diagram, the horizontal branch corresponds to the H-shell and core He-burning phase, during which beta decay is the primary neutrino production process. The steep increase in the neutrino luminosity occurs during the C-burning phase and is driven by thermal neutrino production \citep{odrzywolek2004}. To evaluate the relative significance of various thermal processes, the left panel of Figure \ref{fig:neutrinolum} presents the time evolution of neutrino luminosities from different thermal processes during the C-burning phase for our reference star (30 \msun~and $Z=0.02$ at ZAMS). Additionally, the right panel of Figure \ref{fig:neutrinolum} displays neutrino luminosities as a function of radius for the same star at 210 years after the beginning of C-burning. The figure reveals that the luminosity at the core is predominantly due to electron-positron pair annihilation, although photo-neutrino production is important outside the core. As a consequence, pair annihilation accounts for $\gtrsim75$ \% of the total luminosity, with the fraction increasing in later stages. We observe similar situations for other model stars.

To assess the prospect of detecting neutrinos at Earth, we need their energy spectrum. Given that pair annihilation is the dominant mechanism, we calculate the neutrino spectrum produced only via pair annihilation, following the steps described in \citet{Kato2015, Kato2017}.

In natural units, the neutrino energy spectrum resulting from pair annihilation is given by
\begin{equation}
\begin{aligned}
    \frac{dQ^{\nu,\bar{\nu}}_{N}}{{dE}_{\nu,\bar{\nu}}} = \frac{E_{\nu,\bar{\nu}}}{(2\pi)^2} \iint \frac{E_{\bar{\nu},\nu}}{2(2\pi)^2} R(E_\nu,E_{\bar{\nu}},\cos\Theta) \\
    \times dE_{\bar{\nu},\nu} \, d\cos\Theta,
\end{aligned}
\end{equation} 
where $Q^{\nu,\bar{\nu}}_{N}$ is the number of neutrinos ($\nu$) and anti-neutrinos ($\bar{\nu}$) per unit volume per unit time, and ${E}_{\nu,\bar{\nu}}$ and $\Theta$ are the energy of $\nu$ and $\bar{\nu}$ and the angle between between $\nu$ and $\bar{\nu}$ directions, respectively. In the above equation, $R(E_\nu,E_{\bar{\nu}},\cos\Theta)$ is given by 
\begin{equation}
     R(E_\nu,E_{\bar{\nu}},\cos\Theta)=\frac{8G_F^2}{\left(2\pi\right)^2}\ (\beta_1I_1+\beta_2I_2+\beta_3I_3),
\end{equation}
where $G_F=1.166364\times{10}^{-11}~{\mathrm{MeV}}^{-2}$ is the Fermi coupling constant. Here, $\beta_1=\left(C_V-C_A\right)^2$, $\beta_2=\left(C_V+C_A\right)^2$, and $\beta_3=C_V^2-C_A^2$, and two coupling vectors for electron neutrinos ($\nu_e$) and electron anti-neutrinos ($\bar{\nu_e}$) are $C_V=1/2+{2\times\left(0.226\right)}^2$ and $C_A=1/2$, respectively. For other flavors (i.e., muon and tau neutrinos and anti-neutrinos, $\nu_\mu$, $\bar{\nu_\mu}$, $\nu_\tau$, $\bar{\nu_\tau}$), $C_V=-1/2+{2\times\left(0.226\right)}^2$ and $C_A=-1/2$ are used. For the Fermi integration terms $(I_1,\ I_2,\ I_3)$, refer to \citet{Kato2017}. We note that the neutrino emissivity $Q$ (in units of energy per volume per time) can be calculated as $\int ({dQ^{\nu,\bar{\nu}}_{N}}/{{dE}_{\nu,\bar{\nu}}}) {E}_{\nu,\bar{\nu}} {dE}_{\nu,\bar{\nu}}$. The emissivity calculated in this way is consistent with that obtained using the {\tt MESA} thermal neutrino module, although the model includes contributions from all thermal processes.

\begin{figure}[t]
\vskip -0.2 cm
\hskip -0.4 cm
\includegraphics[width=0.55\textwidth]{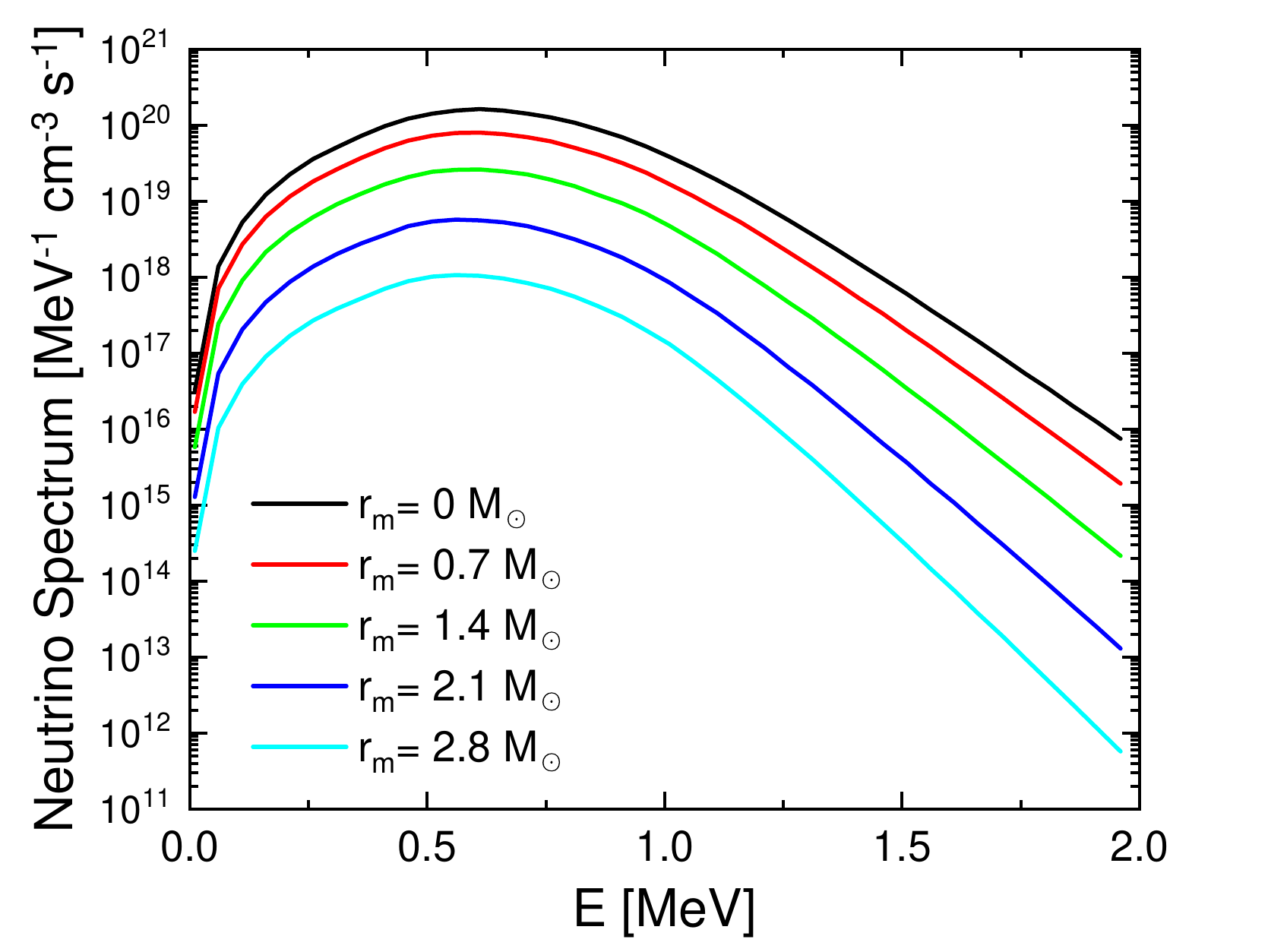}
\vskip -0.2 cm
\caption{Energy spectrum of electron neutrinos from pair annihilation, produced in the reference star at 210 years after the beginning of C-burning. Lines with different colors plot the spectra, in units of number per energy per volume per time, at various radii corresponding to the specified mass coordinate values.}
\label{fig:NeuSpec}
\end{figure}

\begin{figure*}[t]
\vskip 0.2 cm
\hskip -0.2 cm
\begin{tabular}{cc}
\includegraphics[width=0.92\textwidth]{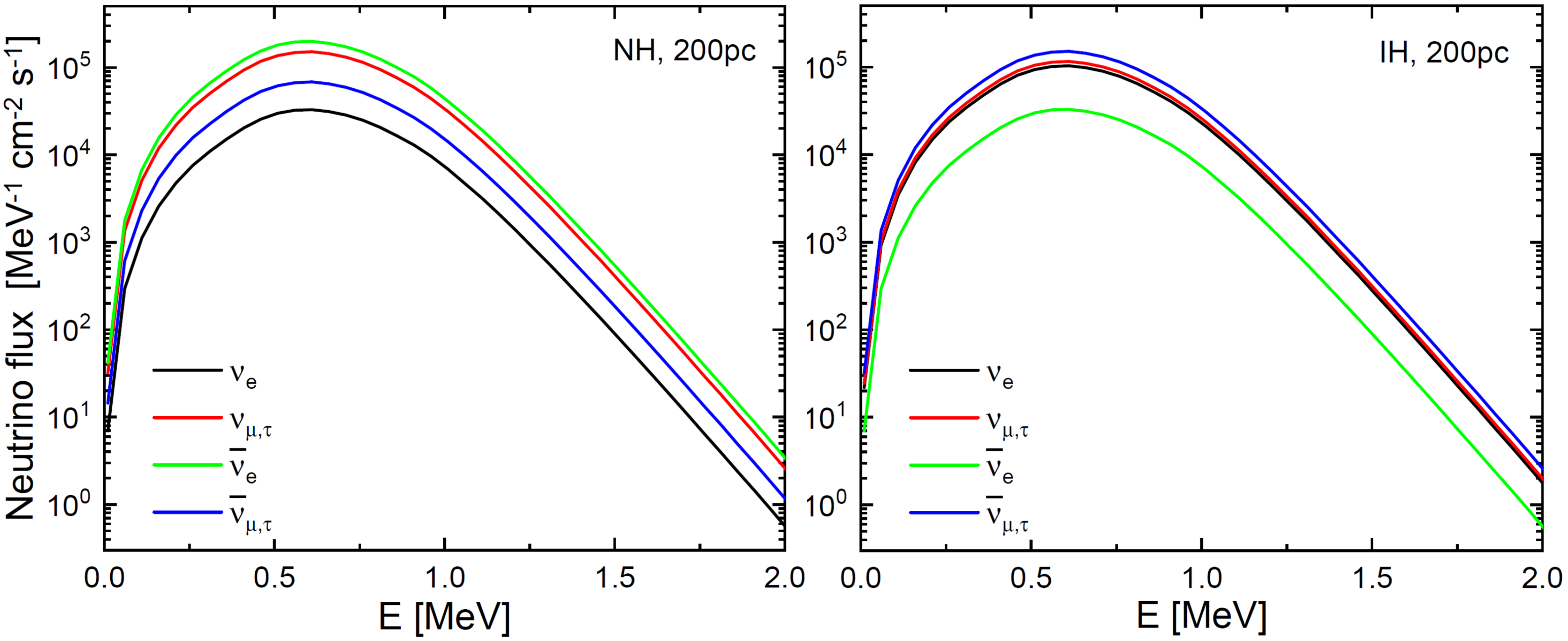}
\end{tabular}
\vskip -0.1 cm
\caption{Fluxes of neutrinos of various flavors at Earth as a function of energy, originating from the reference star at 210 years after the beginning of C-burning, located at a distance of 200 pc. The fluxes incorporating neutrino oscillations for both normal mass ordering (left panel) and inverted mass ordering (right panel) are shown.}
\label{fig:flux_at_earth}
\end{figure*}

Figure \ref{fig:NeuSpec} presents the energy spectrum of $\nu_e$ generated via pair annihilation at various radii, expressed in units of number per energy per volume per time, for the reference star at 210 years after the beginning of C-burning. A nearly equal amount of $\bar{\nu_e}$ is also produced \citep{odrzywolek2004}. On the contrary, the production of $\nu_\mu$, $\bar{\nu_\mu}$, $\nu_\tau$, and $\bar{\nu_\tau}$ is in much smaller quantities. The peak of the spectrum is governed by the gas temperature, and hence shifts to lower energies from the core to the outer regions. Given that neutrino production is largest at the core, the peak energy of the spectrum of all neutrinos produced within the star is primarily determined by the central temperature, and is $\sim0.6$ MeV for this model star.

In estimation of the neutrino flux reaching Earth, the effects of neutrino oscillations enter. For it, we adopt the formulae previously used to estimate neutrino fluxes from pre-supernovae \citep[e.g.,][]{Kato2015, Kato2017, Patton2017b}. These formulae take into account the MSW effect within the star and vacuum oscillations between the star and Earth. Figure \ref{fig:flux_at_earth} shows the fluxes of neutrinos of various flavors, expected at terrestrial detectors; the fluxes are calculated for the reference star at 210 years after the beginning of C-burning, assuming that the star is located 200 pc away, which corresponds to the distance between Betelgeuse and Earth. The fluxes of neutrino flavors depend on neutrino mass ordering, and we consider oscillations for both normal and inverted hierarchies. With normal mass ordering, most neutrinos are converted into $\bar{\nu_e}$ and also $\nu_\mu$ and $\nu_\tau$ by the time they arrive at terrestrial detectors. In contrast, with inverted mass ordering, neutrinos primarily transform into $\nu_e$ and also $\nu_\mu$, $\bar{\nu_\mu}$, $\nu_\tau$, and $\bar{\nu_\tau}$. The figure indicates that regardless of mass ordering, the total flux reaching Earth is on the order of $\sim10^5$ cm$^{-2}$s$^{-1}$; it peaks at $\sim0.6$ MeV, the same as the source spectrum shown in Figure \ref{fig:NeuSpec}. As a matter of fact, $\sim90 \%$ of neutrinos have energies less than 1 MeV.

\section{Detectability of Neutrinos from RSGs}\label{sec3}

\subsection{Hybrid detector for sub-MeV neutrinos}

With energies mostly in the sub-MeV range, neutrinos from C-burning stars shown in Figure \ref{fig:flux_at_earth} pose significant challenges for detection. WCDs, listed in the introduction, have a detection threshold of $\sim 3.5$ MeV \citep{Abe2016}; below this threshold, data are generally not taken due to high background noise from radioactive isotopes. To date, sub-MeV neutrinos have typically been detected in scintillator-based detectors \citep[e.g., Borexino,][]{BorexinoPP, BorexinoCNO}. In these detectors, fluorescence photons are emitted {\it isotropically} when an electron is liberated following an elastic collision with an incoming neutrino; hence, directional information about the incoming neutrino is difficult to be obtained, which limits their applicability for astronomical studies.

To overcome these limitations, researchers have explored methods to enable directional measurement in scintillator-based detectors. For instance, \citet{Mukhopadhyay2020} demonstrated the directional sensitivity of such detectors by utilizing inverse beta decay reactions due to presupernova neutrinos. In this approach, the spatial correlation between the produced positron and neutron provides directional information for anti-neutrinos. Another notable effort is the correlated and integrated directionality (CID) method, which was investigated for measuring the directionality of sub-MeV solar neutrinos in Borexino experiments \citep{Agostini2022PhRvD.105e2002A, Agostini2022PhRvL.128i1803A}. This technique aims to detect Cherenkov photons produced by solar neutrinos in traditional liquid scintillators and correlate their emission with the Sun’s position to infer neutrino directionality.

Another possible approach to obtaining  the directional information for sub-MeV neutrinos would be the use of a hybrid detector that combines Cherenkov and scintillator techniques \citep{Askins2020}. In such a detector, an electron emits both Cherenkov and fluorescence photons. Since the number of fluorescence photons significantly exceeds that of Cherenkov photons, hybrid detectors can achieve a lower detection threshold and higher detection efficiency compared to WCDs in the sub-MeV range. In contrast, Cherenkov photons, which are emitted {\it directionally}, allow for direction measurement. To utilize this advantage, it is necessary to distinguish Cherenkov photons from fluorescence photons within the hybrid detector.

Recently, efforts have been made to develop technologies to separate Cherenkov photons from fluorescence photons \citep{Li2016, Biller2020, Kaptanoglu2022}. The separation relies on the fact that Cherenkov photons are, on average, emitted earlier than fluorescence photons. If photodetectors in a hybrid detector have a high temporal resolution on the order of a hundred picoseconds, it becomes feasible to differentiate Cherenkov photons from fluorescence photons. In fact, advanced photodetectors, such as ultra-fast photomultiplier tubes (PMTs) with timing precision of $\sim25$ picoseconds \citep{Fu2020} and large area picosecond photodetectors (LAPPD) \citep{Lyashenko2020_LAPPD}, show potential for achieving the required temporal resolution.

\begin{figure}[t]
\vskip -0.2 cm
\hskip -0.4 cm
\includegraphics[width=0.55\textwidth]{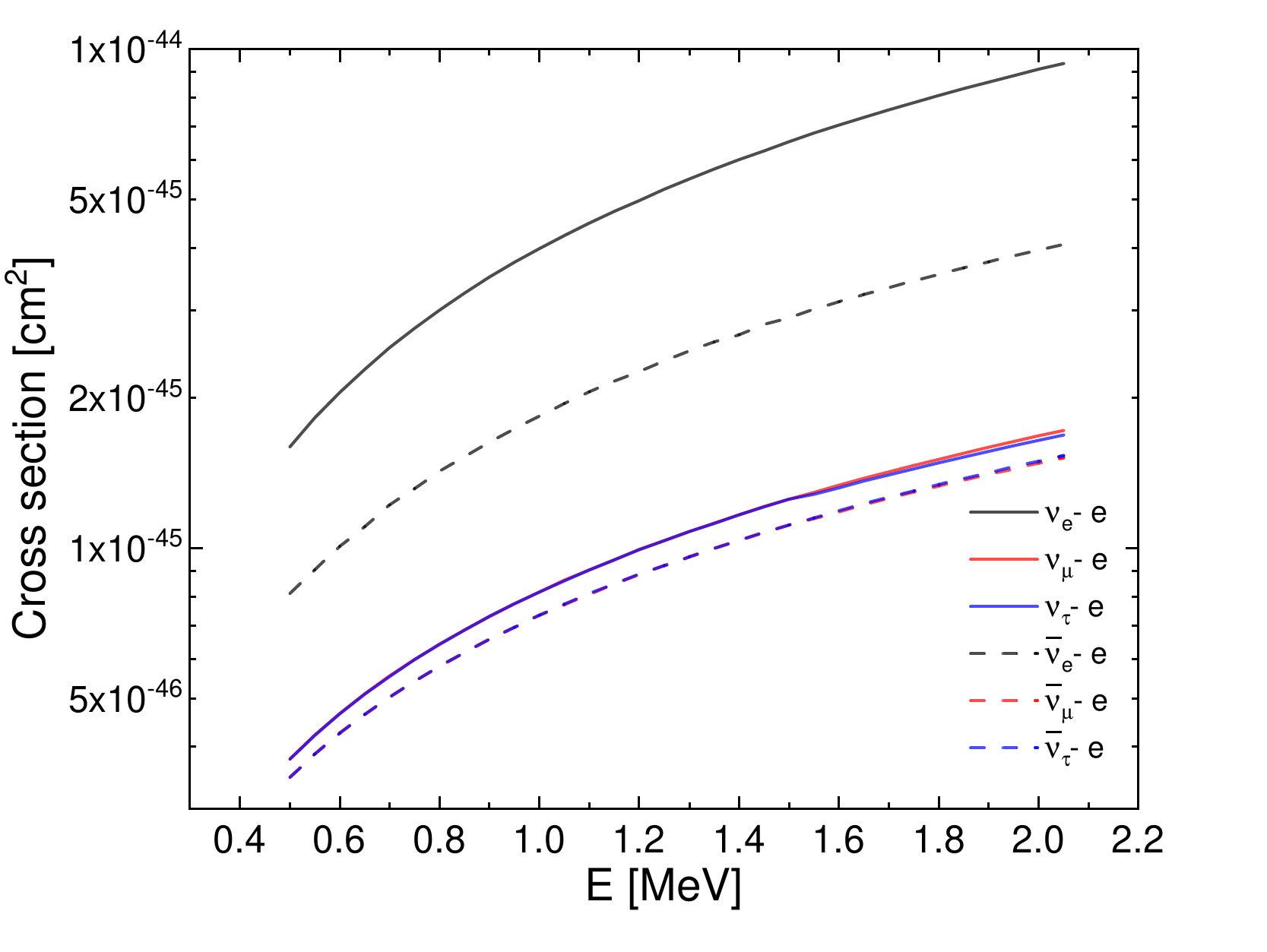}
\vskip -0.3 cm
\caption{Cross sections for elastic scattering between neutrinos of various flavors and electrons as a function of neutrino energy. Red and blue lines nearly overlap for both solid and dashed curves.}
\label{fig:cross-section}
\end{figure}

To assess the performance of hybrid detectors for neutrinos from C-burning RSGs, we run Monte Carlo (MC) simulations using software packages {\tt NuWro} \citep{Golan2012_Nuwro} and {\tt Geant4} \citep{Agostinelli2003, Allison2006, Allison2016}. {\tt NuWro} simulates neutrino interactions with matter in experimental settings. In the sub-MeV range, elastic scattering is the dominant interaction channel of neutrinos. By utilizing the extension for neutrino-electron scattering processes described in \citet{Zhuridov2021_NuWro}, we obtain the cross sections of interactions between all neutrino flavors and electrons (see Figure \ref{fig:cross-section}) and the expected energy and directional distributions of recoiled electrons following scattering. Electron-type neutrinos, $\nu_e$ and $\bar{\nu_e}$, have larger cross sections compared to others, $\nu_\mu$, $\bar{\nu_\mu}$, $\nu_\tau$, and $\bar{\nu_\tau}$.

\begin{figure}[t]
\vskip 0.1 cm
\hskip -0.4 cm
\includegraphics[width=0.5\textwidth]{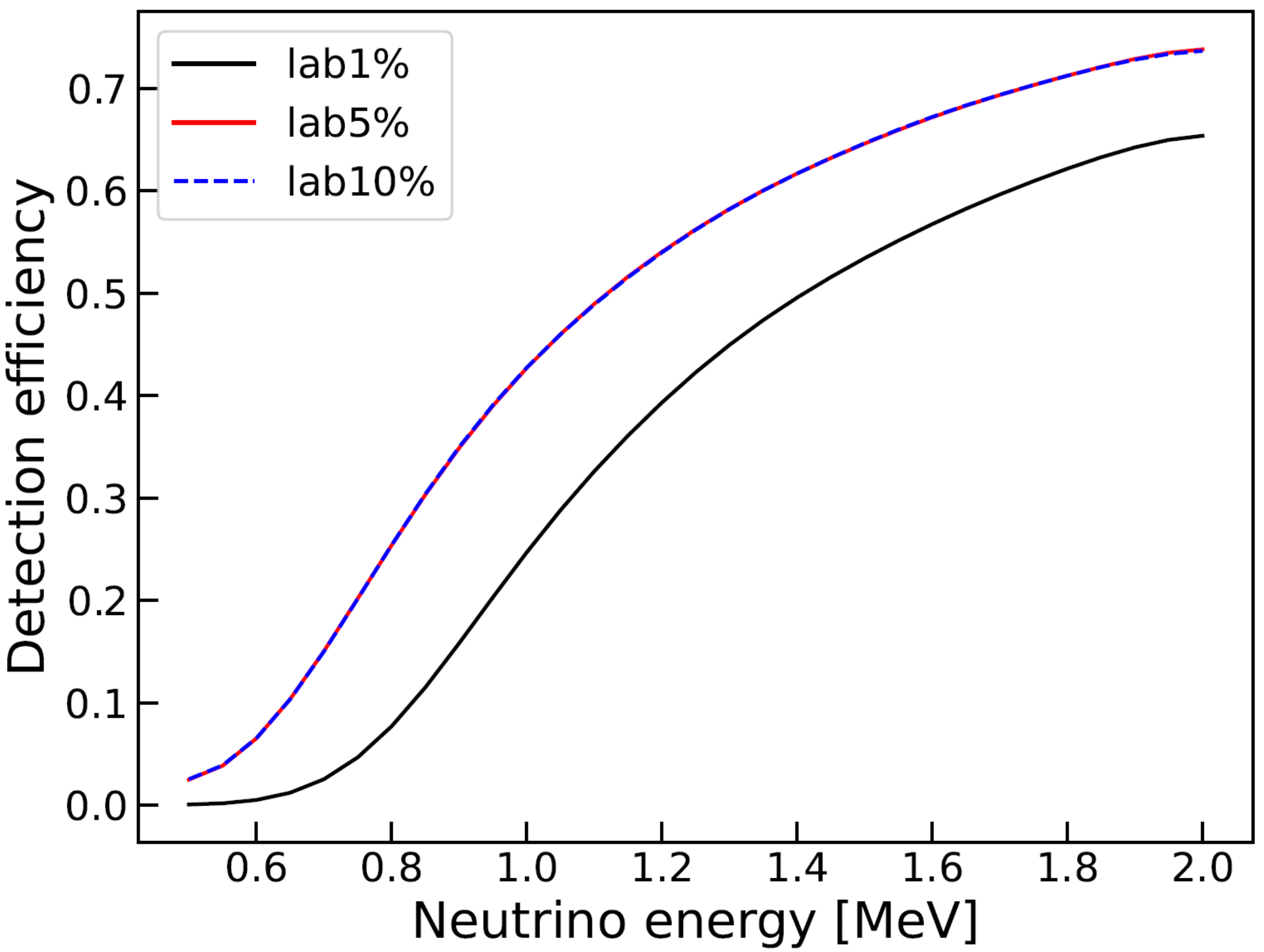}
\vskip -0.1 cm
\caption{Efficiency of detections among interactions between incoming neutrinos and WbLS in our model hybrid detectors. The detectors are filled with a WbLS composed of water and varying concentrations of LAB: 1\% (black solid), 5\% (red solid), and 10\% (blue dashed). Red solid line and blue dashed line nearly overlap.}
\label{fig:deteff}
\end{figure}

{\tt Geant4} can simulate the emission of Cherenkov and fluorescence photons from recoiled electrons, as well as their registration on photodetectors in model detectors. In this study, we consider idealized hybrid detectors with a spherical geometry and a volume of 50 kilotons (kt), comparable to that of Super-Kamiokande, which are filled with a water-based liquid scintillator (WbLS), consisting of water mixed with 1\%, 5\%, and 10\% linear alkyl benzene (LAB). For the energy and directional distributions of recoiled electrons, the output of {\tt NuWro} is used. Photon emission is modeled using experimental data from the CHESS experiment \citep{Caravaca2020_CHESS}. The detector wall is assumed to be fully covered with LAPPDs, which feature a quantum efficiency of $ \gtrsim30\%$, a time resolution of $\sim70$ ps, and a spatial resolution of $\sim1$ mm \citep{Shin2024_LAPPD}.

In our MC simulations, 50,000 interactions between incoming neutrinos and WbLS are generated with neutrino energies ranging from 0.5 to 2 MeV, for each model detector with varying mixtures of LAB. The interaction locations and incoming directions are randomly assigned. For each interaction, we calculate the number of Cherenkov and fluorescence photons that are produced and registered to LAPPDs.

\begin{figure}[t]
\vskip -0.2 cm
\hskip -0.5 cm
\includegraphics[width=0.52\textwidth]{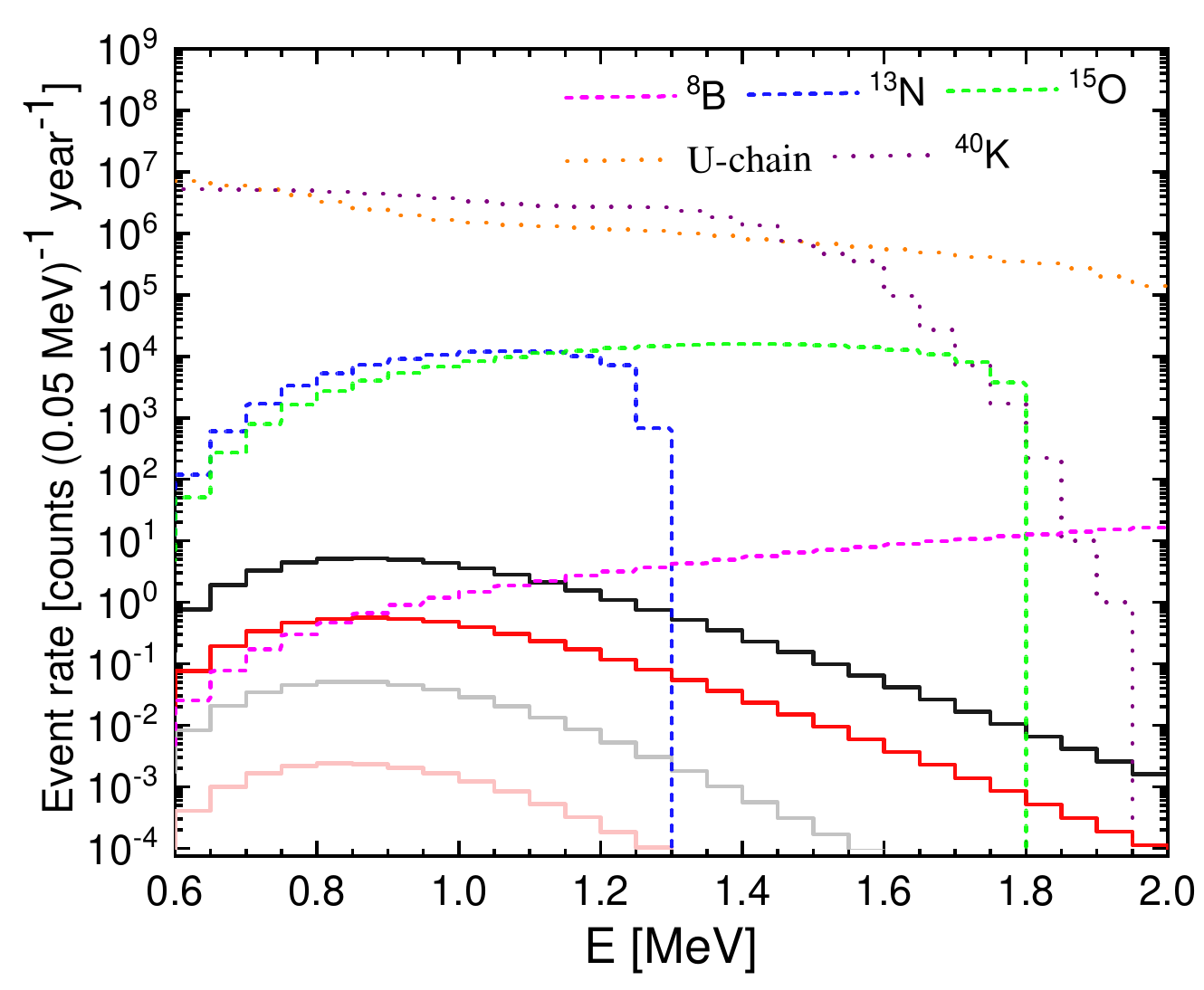}
\vskip -0.2 cm
\caption{Event rates as a function of neutrino energy at a hybrid detector with 5\% LAB. The event rates estimated for neutrinos from two model stars located 200 pc away, initially with 20 \msun~and $Z=0.02$ (red lines) and 30 \msun~and $Z=0.02$ (black lines), are shown at the beginning (light color) and end (dark color) of C-burning. For comparison, the event rates for solar neutrinos from decay of $^{8}$B (magenta dashed line), electron capture on $^{13}$N (blue dashed line) and $^{15}$O (green dashed line) are also shown, along with noise event rates from the decay of $^{40}$K (purple dotted line) and the uranium chain (orange dotted line).}
\label{fig:Eventrate}
\end{figure}

We then evaluate the detection efficiency, defined as the fraction of interactions for which neutrino energy can be reconstructed, using two criteria. First, to account for the inherent dark noise in photodetectors, interactions are assumed to be triggered only when the number of photons detected exceeds a predefined threshold. Based on \citet{Bonventre2018}, $\sim20$ noise hits per kHz are expected during a 200 ns trigger window in a 50 kt detector. Accordingly, we set the trigger threshold to be a minimum of total 50 Cherenkov and fluorescence photons from a recoiled electron being detected by the LAPPDs. Second, since sufficient Cherenkov photons are necessary to extract directional information, we require the production of at least 10 Cherenkov photons. Although these criteria are somewhat arbitrary $-$ and the second criterion can only be applied to Monte Carlo simulations, not real detectors $-$ they would not alter the main conclusions of this work. Figure \ref{fig:deteff} illustrates the detection efficiency in our model hybrid detectors with 1\%, 5\%, and 10\% LAB; the case of 1\% LAB shows a lower detection efficiency compared to the other two cases, while the efficiencies for the cases of 5\% and 10\% LAB are nearly indistinguishable.

\subsection{Event rate} \label{sec:eventrate}

The rate of events detected at our model detectors can be estimated as
\begin{equation}
    r\ ({\rm MeV}^{-1}{s}^{-1})=f\times\sigma\ ({\rm cm}^2)\times\varphi\ ({\rm MeV}^{-1}{\rm cm}^{-2}{s}^{-1})\times N, 
    \label{eq:rates}
\end{equation}
where $f$, $\sigma$, $\varphi$, and $N$ are the detection efficiency, cross section, neutrino flux, and number of target particles in the detector, respectively. Employing $\sigma$ and $f$ from Figures \ref{fig:cross-section} and \ref{fig:deteff}, along with $N$ for 50 kt, we calculate $r$ using $\varphi$ for model stars located 200 pc away (such as those in Figure \ref{fig:flux_at_earth}).

Figure \ref{fig:Eventrate} presents the resulting event rates as a function of neutrino energy at a detector with 5\% LAB; the results are shown for two model stars with initially 20 \msun~and 30 \msun~(both with $Z=0.02$), at the beginning and end of C-burning. The total number of events integrated over all energies, detected over one-year period, is estimated to range from $\sim0.02$ to $\sim50$ for the cases shown in the figure. We point out that the event rates in the figure are calculated assuming normal mass ordering, where $\bar{\nu_e}$ is the dominant neutrino flavor to be detected. The event rates with inverted mass ordering turn out to be nearly identical. With inverted mass ordering, the most important flavor is $\nu_e$; $\varphi$ of $\nu_e$ for inverted mass ordering is smaller than that of $\bar{\nu_e}$ for normal mass ordering (see Figure \ref{fig:flux_at_earth}), but the cross section of $\nu_e$ is higher (see Figure \ref{fig:cross-section}). Although not shown here, other model stars exhibit generally similar values, with the total event rates spanning $\sim0.01 - 50$ per year for the entire set of model stars.

\begin{figure}[t]
\vskip -0.3 cm
\hskip -0.3 cm
\includegraphics[width=0.54\textwidth]{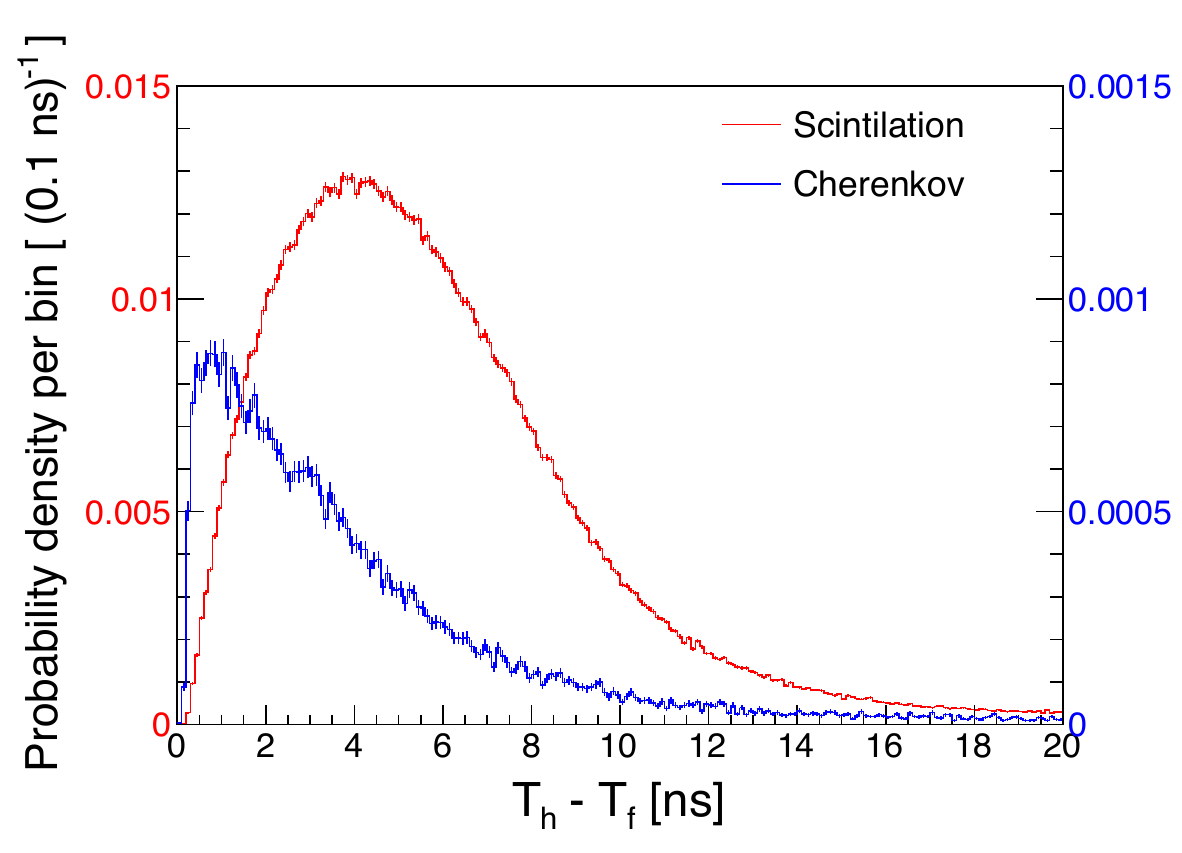}
\vskip -0.3 cm
\caption{Cherenkov (blue line) and fluorescence (red line) photons, produced by recoiled electrons following interactions of 1.0 MeV electron neutrinos with electrons located randomly within a model detector with 5\% LAB, as a function of $T_h-T_f$ (see the text for the definition of $T_f$ and $T_h$). The plotted quantity is the number of photons per bin in MC simulations, normalized by the total number of photons. Note that the $y$-axis scale for fluorescence photons is ten times larger than that for Cherenkov photons.}
\label{fig:timing_hist}
\end{figure}

For comparison, Figure \ref{fig:Eventrate} illustrates the event rates for solar CNO neutrinos, specifically those from electron capture on $^{13}$N (blue solid line) and $^{15}$O (green solid line) \citep{Bahcall1988_solarneu}, at the same model detector. The figure also includes event rates from background noise sources, including the decay of $^{40}$K (purple dotted line) and the uranium chain (orange dotted line). While detailed descriptions of these noise sources are provided in \citet{Bonventre2018}, the event rates are derived by scaling the values in \citet{Askins2020} to match the configuration of our model detector. We see that the event rates of solar CNO neutrinos exceed those of C-burning star neutrinos by at least a factor of $10^3$. Moreover, the noise event rates are larger by another factor of $\sim 10^3$, making them at least $\sim 10^6$ times larger compared to those of C-burning star neutrinos. Given a signal-to-noise ratio $\lesssim10^{-6}$, getting a {\it signature} of neutrinos from nearby C-burning RSGs seems to be challenging with the current model detectors.
 
\subsection{Angular resolution}\label{sec:direction}

\begin{figure}[t]
\vskip -0.3 cm
\hskip -0.4 cm
\includegraphics[width=0.55\textwidth]{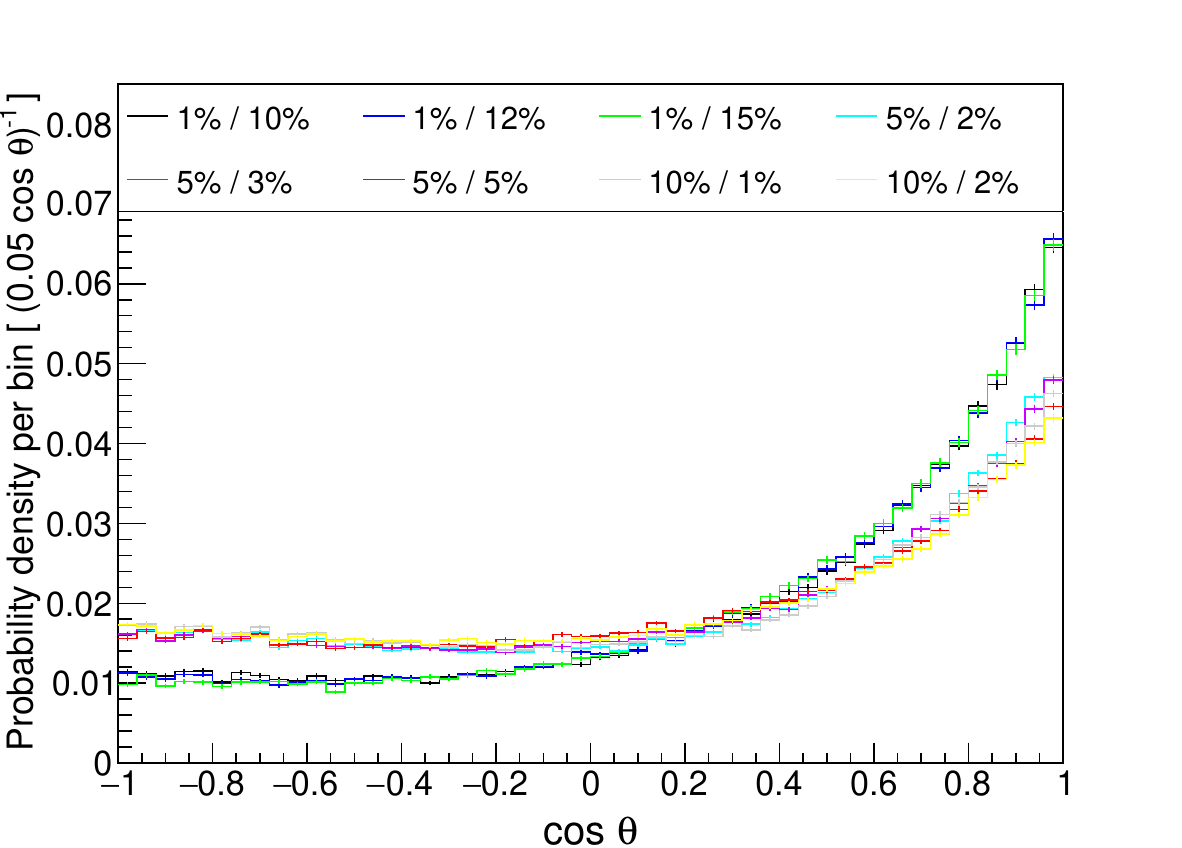}
\vskip -0.1 cm
\caption{Probability distribution of reconstructed directions as a function of $\theta$, the angle between the true incoming direction of electrons and the reconstructed direction, for model detectors with varying LAB fractions and selection cuts. See the text for the definition of selection cut. The plotted quantity is the number of events per bin in MC simulations, normalized by the total number of events.}
\label{fig:directional_acc}
\end{figure}

We note that noise events are isotropic, whereas astrophysical events are directional, pointing toward their sources. Hence, measuring direction would enhance the ability to separate astrophysical events from noises. As a straightforward estimate, if the direction of an event is confined within a solid angle $\Omega$ around a known astrophysical source, the rates of noise events can be effectively reduced by a factor of ${\Omega}/{4\pi}$. 
Direction reconstruction techniques for detectors using WbLS are still under discussion, especially for the sub-MeV range \citep[see][and references therein]{Askins2020}. Below, we present an analysis of angular resolution for our model detectors using MC simulation results.

As noted above, Cherenkov photons are, on average, produced earlier than fluorescence photons. After production, these photons propagate through WbLS before hitting photodetectors inside detectors. We denote the flight time of photons as $T_f$, and define the hit time at photodetectors, since electrons are liberated, as $T_h$. In real detector experiments, $T_h$ is recorded directly. On the other hand, $T_f$ is estimated by reconstructing the interaction locations using photon rings \citep[e.g.,][]{Allega2024A}. With our MC simulations, $T_h$ is extracted from the simulation data, while $T_f$ is calculated using the known interaction locations. The difference, $T_h-T_f$, is essentially the photon production time relative to the liberation of electrons. Figure \ref{fig:timing_hist} presents the timing distributions of Cherenkov and fluorescence photons produced by recoiled electrons as a function of $T_h-T_f$ in our model detector with 5 \% LAB. The plot confirms that Cherenkov photons start to be produced almost instantaneously following the interaction, whereas the production of fluorescence photons gradually increases over time. As a result, Cherenkov photons dominate the initial production, but after $\sim0.3$ ns, the number of fluorescence photons surpasses that of Cherenkov photons.

Figure \ref{fig:timing_hist} suggests that Cherenkov photons can potentially be separated from fluorescence photons based on their timing distributions and used to reconstruct the direction of incoming electrons. However, there is no definitive cut in $T_h-T_f$ that guarantees the best reconstruction. Selecting a smaller $T_h-T_f$ increases the fraction of Cherenkov photons but reduces their total number. Conversely, a large $T_h-T_f$ results in a higher photon number but a lower fraction of Cherenkov photons. To find an optimal trade-off, we conduct tests for our hybrid detectors using MC simulation data. Two parameters are evaluated: the selection cut, defined as the fraction of earliest produced photons, and the mixture of LAB, which affects the accuracy of direction reconstruction. For instance, a $5\%$ selection cut implies using the $5\%$ of photons with the smallest $T_h-T_f$ for reconstruction. The direction is then determined by averaging the vectors connecting the interaction location to the positions where these photons are registered on photodetectors.

The results of direction reconstruction are presented in Figure \ref{fig:directional_acc} and Table \ref{tab:ang_acc}, illustrating the consequences of varying LAB fractions and selection cuts. Here, $\theta$ is the angle between the true incoming direction of elections and the reconstructed direction. The figure plots the probability distribution of reconstructed directions as a function of $\theta$, while the table lists the fraction of reconstructions with $\theta$ less than specific values. Smaller LAB fractions result in better angular reconstruction, but it comes at the cost of reduced detection efficiency, as shown in Figure \ref{fig:deteff}. The optimal selection cut for maximizing performance depends on the LAB fraction. For our model detectors, we pick up $\sim5\%$ LAB and $\sim2\%$ selection cut as the optimal combination. With these, $\sim28\%$ of events are reconstructed within $\leq45^\circ$ and $\sim43\%$ within $\leq60^\circ$. Thus, we regard that in our model detectors, the angular resolution for detecting sub-MeV neutrinos is on the order of tens of degrees. The solid angle for a cone with $45^\circ$ radius corresponds to ${\Omega}/{4\pi}\approx0.15$. Hence, the directional information should be helpful, but seems insufficient to overcome the low signal-to-noise ratio in the effort to detect neutrinos from C-burning RSGs.

\begin{table}[t]
{\centering
\caption{Fraction of reconstructions with $\theta$ less than specific values}
\label{tab:ang_acc}
\begin{tabular}{ccccc}
\toprule
LAB & Selection & $\theta$       & $\theta$       & $\theta$       \\
    & cut       & $\leq30^\circ$ & $\leq45^\circ$ & $\leq60^\circ$ \\
\hline
1\%   & 10\% & 0.223  & 0.345  & 0.518 \\
1\%   & 12\% & 0.224  & 0.346  & 0.520 \\
1\%   & 15\% & 0.223  & 0.345  & 0.520 \\
5\%   & 2\%  & 0.175  & 0.276  & 0.427 \\
5\%   & 3\%  & 0.170  & 0.268  & 0.417 \\
5\%   & 5\%  & 0.158  & 0.253  & 0.402 \\
10\%  & 1\%  & 0.166  & 0.264  & 0.413 \\
10\%  & 2\%  & 0.156  & 0.249  & 0.380 \\
\hline
\end{tabular}}
\end{table}

\section{Summary and Discussion} \label{sec:discussion}

In this paper, we evaluated the flux of neutrinos that are emitted from evolved massive stars, particularly RSGs in the C-burning phase, and reach Earth. Since the C-burning phase lasts a few hundred to a thousand years, such stars can be considered continuous neutrino sources like the Sun, rather than transient neutrino sources such as supernovae. Expecting several C-burning RSGs in our Galaxy, they could be viable targets for MeV neutrino astronomy. We then assessed the possibility of detecting these neutrinos at terrestrial detectors, focusing on idealized hybrid detectors filled with WbLS and equipped with LAPPD.

Using the {\tt MESA} code, we followed the evolution from ZAMS to the end of core C-burning for a set of model stars with different initial masses, 12, 15, 20, 25, 30, and 35 \msun, and different metallicities, $Z =$ 0.01, 0.015, and 0.02. During C-burning, a significant amount of thermal neutrinos are produced, primarily via pair annihilation. The peak of the neutrino energy spectrum is determined by the core temperature and appears at sub-MeV, around $\sim0.6$ MeV. Assuming that model stars are located 200 pc from Earth, corresponding to the distance to Betelgeuse, the nearest RSG, we obtained the neutrino fluxes reaching Earth as high as $\sim10^5$ cm$^{-2}$s$^{-1}$.

With these fluxes, we examined the detection of neutrinos at model hybrid detectors with a 50 kt volume containing $1 - 10\%$ of LAB. The event rate is estimated to range $\sim0.01-50$ per year, with the highest rate occurring during the final stage of carbon burning. Using the fact that Cherenkov photons are emitted earlier than fluorescence photons and are directional, the angular resolution of detected events is estimated to be on the order of tens of degrees.

While the event rate as large as 50 per year may be considered fair, it is $\sim10^3$ times smaller than that for solar neutrinos from electron capture on $^{13}$N and $^{15}$O. Furthermore, it is $\sim10^6$ times smaller than the event rate due to background noise sources, such as the decay of $^{40}$K and the uranium chain. Combined with the rather poor angular resolution, it would be challenging to get a clear signature of neutrinos from carbon-burning RSGs at detectors like our model detectors. However, we point out that detector technologies keep improving. For instance, in liquid scintillator blended with bismuth compounds, the emission of fluorescence photons is delayed \citep[e.g.,][]{ren2024}. Hence, in detectors using it, the separation of Cherenkov and fluorescence photons would be easier, potentially the angular resolution being improved. Thus, although detecting neutrinos from carbon-burning RSGs seems difficult with currently available detector technologies, it could become feasible in the future as these technologies continue to advance.

Finally, we note that, during the lifetime of massive stars prior to SN explosion, most neutrinos are emitted at energies below $\sim3$ MeV \citep{Farag2024}. Therefore, hybrid detectors capable of providing observational windows for MeV neutrinos would become a valuable tool for neutrino astronomy, and this work demonstrates its potential. As a matter of fact, more promising targets for such detectors should be stars in the O and Si-burning stages, although stars in these stages would have a small chance of being found in our Galaxy or the Local Group due to their short lifespans, as mentioned in the introduction. The potential for detecting neutrinos from these stars at WCDs with a threshold of $\sim 3.5$ MeV and providing SN early warning has been previously explored (see references in the introduction). However, given that most of such neutrinos have energies below $\sim3$ MeV, it would be interesting to investigate how effectively they can be detected in hybrid detectors. We leave it as a topic for future research.

\begin{acknowledgments}
This work was supported by the National Research Foundation (NRF) of Korea through grants 2020R1A2C2102800 and 2022R1F1A1073890.
\end{acknowledgments}

\bibliography{ref1_v2}
\bibliographystyle{aasjournal}

\end{document}